\pdfoutput=1

\documentclass[11pt]{article}

\usepackage[final]{acl}

\usepackage{times}
\usepackage{latexsym}
\usepackage{colortbl}
\usepackage{xcolor}
\usepackage{array}
\usepackage{tabularx}
\usepackage{float}

\usepackage[T1]{fontenc}

\usepackage[utf8]{inputenc}

\usepackage{microtype}

\usepackage{inconsolata}
\usepackage{longtable}
\usepackage{graphicx}
\usepackage{amsmath}
\usepackage{multirow}
\usepackage{booktabs} 
\title{Decomposition Dilemmas: Does Claim Decomposition Boost or Burden Fact-Checking Performance?}

\author{Qisheng Hu \quad Quanyu Long \quad Wenya Wang \\
        Nanyang Technological University \\
        \texttt{qisheng001@e.ntu.edu.sg, quanyu001@e.ntu.edu.sg, wangwy@ntu.edu.sg}
        }

\begin{document}
\maketitle
\begin{abstract}
Fact-checking pipelines increasingly adopt the \textit{Decompose-Then-Verify} paradigm, where texts are broken down into smaller claims for individual verification and subsequently combined for a veracity decision. While decomposition is widely-adopted in such pipelines, its effects on final fact-checking performance remain underexplored. Some studies have reported improvements from decompostition, while others have observed performance declines, indicating its inconsistent impact. To date, no comprehensive analysis has been conducted to understand this variability. To address this gap, we present an in-depth analysis that explicitly examines the impact of decomposition on downstream verification performance. Through error case inspection and experiments, we introduce a categorization of decomposition errors and reveal a trade-off between accuracy gains and the noise introduced through decomposition. Our analysis provides new insights into understanding current system's instability and offers guidance for future studies toward improving claim decomposition in fact-checking pipelines.
\footnote{Source code available at \url{https://github.com/qishenghu/Decomp_Dilemmas}}
\end{abstract}

\section{Introduction}

\begin{figure*}[ht!]
    \setlength{\abovecaptionskip}{0.2cm}
    \setlength{\belowcaptionskip}{-0.4cm}
     \centering
         \includegraphics[width=0.98\textwidth]{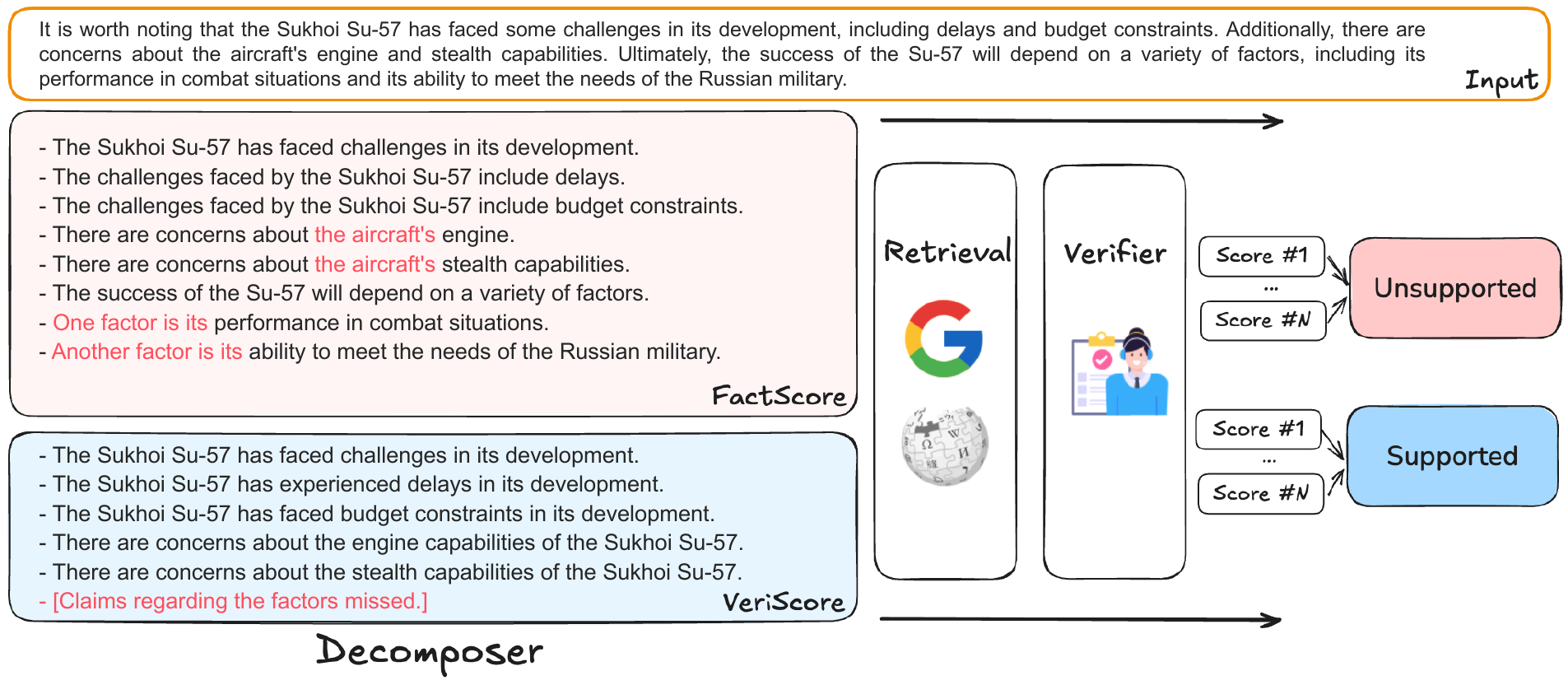}
        \caption{An overview of the \textit{Decompose-Then-Verify} pipeline employed in this study, which comprises four key stages: decomposition, retrieval, verification, and aggregation of sub-claim results. This figure illustrates how different decomposition methods, such as FactScore~\citep{min-etal-2023-factscore} and VeriScore~\citep{song2024veriscore}, can lead to divergent decomposing outcomes. In this example, FactScore generates ambiguous sub-claims, while VeriScore omits key information (e.g., ``Ultimately, the success of the Su-57...'') from the input.
        }
         \label{fig:pipline}
\end{figure*}

Fact-checking is a critical task that typically involves evaluating the veracity of claims or reports. With the rise of large language models (LLMs), the scope of fact-checking task has expanded to include verifying content generated by LLMs~\citep{sun2024trustllm}. While progress has been made in reducing LLM hallucinations, the problem has not yet been resolved, presenting ongoing challenges in ensuring the factuality of LLM outputs~\citep{huang2023survey}. At the same time, many studies focus on LLM-driven automated fact-checking pipelines~\citep{min-etal-2023-factscore, wei2024long, chern2023factool}, aiming to improve the efficiency of the fact-checking processes.

A common design pattern in these LLM-driven pipelines is the \textit{Decompose-Then-Verify} paradigm, as adopted in frameworks like FactScore~\citep{min-etal-2023-factscore}, FacTool~\citep{chern2023factool} and VeriScore~\citep{song2024veriscore}. It involves decomposing input text into sub-claims (\emph{Decompose}), retrieving supporting information (e.g., Wikipedia, Google Search) for each sub-claim, and using a verifier model to assess the veracity of each sub-claim (\emph{Verify}). The results are then aggregated to produce a final verification. By breaking down complex inputs (decomposition), these pipelines are effective in pinpointing misinformation, enabling a nuanced determination of whether a text is supported, unsupported~\citep{zhu-etal-2023-explain, zhao2024pacar, li-etal-2024-self}, or assigned a quantified score~\citep{min-etal-2023-factscore, wei2024long}.

However, most existing studies primarily focus on the design of fact-checking pipeline architectures. While decomposition is commonly employed in these fact-checking frameworks, the reliability of the decomposition process itself remains insufficiently investigated. Some studies show that the final predictions of these pipelines are sensitive to the decomposition outcomes~\citep{jiang2024core, wanner-etal-2024-closer}. Additionally, it has been observed that a given decomposer does not consistently lead to performance improvements when paired with different verifiers. For example, ~\citet{kamoi-etal-2023-wice} report improvements using a pre-trained Natural Language Inference (NLI) model as the verifier. However, an ablation study by Minicheck~\citep{tang-etal-2024-minicheck}, which employs the same dataset and decomposition method but different NLI models, does not observe consistent benefits from decomposition. Similar inconsistencies are observed in FELM~\citep{zhao2024felm}, which assesses the effect of decomposing responses into segments and further into claims. The study reveals distinct differences between ChatGPT~\citep{chatgptreport} and GPT-4~\citep{gpt4report} as verifiers. Specifically, ChatGPT's accuracy improves with claim-level decomposition, while GPT-4's performance declines.

Despite observed inconsistencies, the underlying reasons remain unclear, and there is limited in-depth analysis of how decomposition affects downstream performance. In this study, we investigate the underlying causes of performance variability by addressing three key questions. First, \textbf{what determines decomposition's effect on fact-checking performance?} Through experiments, we present performance variability across factors such as input complexity, decomposition method design, and verifier strength.  Second, \textbf{what errors may decomposition introduce?} We categorize decomposition error types, analyze their distribution, and validate their usefulness through error reflection. Third, \textbf{what explains the variability in fact-checking performance?} Our analysis reveals a trade-off between the accuracy gains by decomposing inputs into manageable sub-claims and the noise introduced by retrieval and decomposition as the number of sub-claims increases. This trade-off offers an explanation for the observed variability and provides insights for guiding future fact-checking pipeline design.

Here are some key takeaways:
\begin{itemize}
    \item Current popular LLM-driven decomposition methods struggle to consistently improve fact-checking performance across varying input granularities and verifier strengths.
    \item Decomposition methods tend to introduce different errors depending on the objectives—prioritizing high atomicity may over-fragment simple facts, causing redundancy and ambiguity, while emphasizing verifiability can omit essential details.
    \item Decomposition improves performance on simpler sub-claims by reducing complexity, notably benefiting weaker verifiers. However, for stronger verifiers, the marginal accuracy gain from decomposition may not counterbalance the increased noise.
    \item Decomposition can improve the handling of complex inputs; however, while increasing the number of sub-claims may initially enhance performance, the additional noise introduced will gradually offset these gains, eventually leading to performance degradation.
\end{itemize}

\section{Related Work}

\subsection{\textit{Decompose-Then-Verify}}
\textit{Decompose-Then-Verify} pipeline has shown effective for pinpointing errors and improving factual precision~\citep{wang2024openfactcheck}. It involves decomposing text into sub-claims, retrieving supporting materials from a knowledge source, and using a verifier to assess each sub-claim. The individual results are subsequently aggregated to produce a final verification. FactScore~\citep{min-etal-2023-factscore} and SAFE~\citep{wei2024long} proposed decomposing LLM's generated biography into atomic facts and individually verified to report a precision score. FacTool~\citep{chern2023factool} classified the response factuality(`True' or `False') and expands the research scope to include response from more domains (i.e., Knowledge-QA, Math, etc).

\subsection{Decomposition}
\label{sec:decomp_methods}
Claim decomposition has become a widely used technique in fact-checking research. ProgramFC~\citep{pan-etal-2023-fact} introduced a program-guided approach to this technique. QABriefs~\citep{fan-etal-2020-generating} leverages claim decomposition to generate briefs that help crowdworkers improve fact-checking accuracy. Moreover, transforming claim verification into question-like formats~\citep{chen-etal-2024-complex, ousidhoum-etal-2022-varifocal} has proven effective in enhancing fact-checking performance.

To the best of our knowledge, many pipelines rely on LLMs for generating sub-claims using in-context learning.  Both WICE~\citep{kamoi-etal-2023-wice} and FactScore~\citep{min-etal-2023-factscore} design prompts to decompose input into atomic facts, with FactScore further utilizing retrieval to select demonstrations for few-shot prompting. VeriScore~\citep{song2024veriscore}, a variant of FactScore, incorporates a moving-window mechanism and aims to extract only verifiable claims. In addition to prompt-based methods, CORE~\citep{jiang2024core} employs a post-processing approach based on mutual information to retain only the key set of sub-claims.

\begin{table*}[t]
    \centering
    \setlength{\abovecaptionskip}{0.2cm}
    \setlength{\belowcaptionskip}{-0.4cm}
    \small
    \begin{tabular}{cccccc}
        \toprule
        \textbf{Granularity} & \textbf{Dataset} & \textbf{Length} & \textbf{KS} & \textbf{Decomp Method} & \textbf{Verifier} \\
        \midrule
        \multirow{2}{*}{Claim-level} & WICE & 33.65 & Wiki & \multirow{2}{*}{VeriScore/Wice/FactScore} & AlignScore/Minicheck \\
        & CLAIMDECOMP & 35.82 & Google &   & AlignScore/Minicheck/4o-mini \\
        \midrule
        \multirow{2}{*}{Response-level} & FELM & 66.74 & Google & \multirow{2}{*}{VeriScore/Wice/FactScore} & AlignScore/Minicheck/4o-mini \\
        & BINGCHAT & 351.9 & Google &   & AlignScore/Minicheck/4o-mini \\
        \bottomrule
    \end{tabular}
    \caption{Overview of settings in our experiments, including granularity, dataset, average length in tokens, knowledge source, decomposition methods, and verifiers. For experiments on WICE, GPT-4o-mini verifier is excluded because its few-shot prompt from VeriScore is designed for Google Search results, while WICE retrieves in Wiki articles.}
    \label{tab:datasets_overview}
\end{table*}

\section{\textit{Decompose-Then-Verify} Framework Setup}

\subsection{Fact-Checking Task \& Metrics}
In the \textit{Decompose-Then-Verify} pipeline, fact-checking is framed as a binary classification task, aiming to predict whether an input text is `Supported' or `Unsupported' based on retrieved evidence. Following prior work~\citep{laban-etal-2022-summac, tang-etal-2024-minicheck, zhao2024felm}, we use balanced accuracy (BAcc) and F1-score as primary metrics to address potential class imbalance.

\subsection{Pipeline Structure and Settings}
As shown in Figure~\ref{fig:pipline}, we conduct experiments using a \textit{Decompose-Then-Verify} pipeline consisting of four stages: 

\paragraph{Decompose} Given an input text, it is first decomposed into sub-claims using a specified method and language model\footnote{Unless otherwise specified, GPT-4o-mini is used for decomposition by default.}. Language models are commonly used for decomposition with specific instructions and few-shot demonstrations, where the instructions vary based on the decomposition objective. For instance, FactScore emphasizes breaking inputs into atomic facts, while VeriScore focuses on verifiability. As shown in Table~\ref{tab:datasets_overview}, we use the decomposition modules from \textbf{VeriScore}, \textbf{FactScore}, and \textbf{Wice}\footnote{\citet{kamoi-etal-2023-wice} proposed the WICE dataset and the claim-split method for decomposition. We use 'WICE' to refer to the dataset and 'Wice' to refer to the method.} for our experiments. The prompt instructions for these decomposition methods are provided in the Appendix~\ref{appendix:prompt_decomp_methods}. For the baseline, no decomposition is applied, and the original input is directly used for retrieval and verification.

\paragraph{Retrieve} Each sub-claim is used to retrieve evidence from a knowledge source. Following prior work~\citep{li-etal-2024-self, chern2023factool}, we use the Wikipedia corpus for WICE, retrieving the top 3 articles as evidence, and Google Search\footnote{\url{https://serper.dev/}} as the external knowledge source for other datasets, retrieving the top 10 search results as evidence.

\paragraph{Verify} 
The verification process is typically formulated as a natural language inference (NLI) problem, where the claim is treated as the hypothesis and the retrieved evidence as the premise. A language model is then used to determine whether the premise entails the hypothesis (\emph{Supported}) or not (\emph{Unsupported}). To investigate the effect of decomposition when using distinct verifiers, we employ three different models: \textbf{AlignScore}-large~\citep{zha-etal-2023-alignscore}, \textbf{Minicheck} (Bespoke-MiniCheck-7B)~\citep{tang-etal-2024-minicheck}, and the few-shot classification module from VeriScore~\citep{song2024veriscore}, built on \textbf{GPT-4o-mini}~\citep{gpt4ominireport}. AlignScore and Minicheck generate entailment scores ranging from 0 to 1 for each sub-claim and the corresponding retrieved evidence, while GPT-4o-mini's entailment score is computed by exponentiating the log probability of the token `Supported'. More details are provided in the Appendix~\ref{appendix:verifier_details}.

 \paragraph{Aggregate} To reach a final decision, the verification scores of all sub-claims are aggregated. Using the minimum score can be sensitive to verifier errors~\citep{kamoi-etal-2023-wice}. To mitigate this, we follow WICE and adopt the harmonic mean to aggregate the sub-claim scores into a final score. A final score above 0.5 is classified as `Supported', while scores below are labeled `Unsupported'.

\subsection{Datasets}
We explore two dataset granularities: claim-level and response-level. Claim-level datasets contain individual statements for verification, while response-level datasets comprise longer, LLM-generated responses. The rationale for selecting the below datasets is to experiment with diverse text domains and input lengths, all of which provide fine-grained annotations. For the response-level datasets, we perform decontextualization~\citep{choi-etal-2021-decontextualization} with GPT-4o~\citep{gpt4oreport} to ensure that each response is interpretable as standalone text, independent of the original question. More details on dataset processing can be found in Appendix~\ref{appendix:dataset_proc}.

\paragraph{Claim-level Datasets} \mbox{} \\
\noindent\textbf{WICE~\citep{kamoi-etal-2023-wice}} WICE contains natural claims extracted from Wikipedia sections. We use the test data for our experiments.

\noindent\textbf{CLAIMDECOMP~\citep{chen-etal-2022-generating}} it consists of political claims extracted from PolitiFact\footnote{\url{https://www.politifact.com}}. We use a combination of the dev and test sets for our experiments.

\paragraph{Response-level Datasets}\mbox{} \\
\noindent\textbf{FELM~\citep{zhao2024felm}} FELM provides factuality annotations for ChatGPT responses across multiple domains. We use the `World Knowledge' subset for our experiments.

\noindent\textbf{BINGCHAT~\citep{li-etal-2024-self}} BINGCHAT contains factuality annotations for Bing Chat (now Microsoft Copilot) responses across a range of topics.

\definecolor{lightred}{RGB}{255, 204, 204}
\definecolor{lightgreen}{RGB}{204, 255, 204}
\definecolor{lightblue}{RGB}{204, 229, 255}

\section{What Determines Decomposition's Effect on Fact-Checking Performance?}
\label{sec:performance_var}
Based on the inconsistent impact of decomposition observed in the \emph{Decompose-Then-Verify} pipeline, we identify three primary factors contributing to the inconsistencies: \textbf{decomposition method}, \textbf{verifier strength}, and \textbf{input granularity}. Our analysis demonstrates that each factor exerts a distinct influence, with no single method uniformly enhancing downstream performance across all granularities and verifiers. Notably, verifier strength proves to be a key determinant of performance.

\begin{table}[t]
    \centering
    \setlength{\abovecaptionskip}{0.2cm}
    \setlength{\belowcaptionskip}{-0.2cm}
    \small
    \begin{tabular}{llccc}
        \toprule
        \textbf{Verifier} & \textbf{Decomp} & \textbf{Claims} & \textbf{BAcc} & \textbf{F1} \\
        \midrule
        \multirow{4}{*}{AlignScore} 
        & Baseline & 1.00 & 54.80 & 40.78 \\
        & VeriScore & 2.63 & \cellcolor{lightred}52.17 & \cellcolor{lightred}38.81 \\
        & Wice & 3.10 & \cellcolor{lightblue}56.26 & \cellcolor{lightblue}42.23 \\
        & FactScore & 3.91 & \cellcolor{lightblue}56.87 & \cellcolor{lightblue}42.74 \\
        \midrule
        \multirow{4}{*}{Minicheck} 
        & Baseline & 1.00 & 80.01 & 72.32 \\
        & VeriScore & 2.62 & \cellcolor{lightred}74.74 & \cellcolor{lightred}65.16 \\
        & Wice & 3.10 & \cellcolor{lightred}76.81 & \cellcolor{lightred}68.22 \\
        & FactScore & 3.91 & \cellcolor{lightred}71.11 & \cellcolor{lightred}59.90 \\
        \bottomrule
    \end{tabular}
    \caption{Performance and average number of sub-claims after decomposition for different methods on WICE.}
    \label{tab:wice}
\end{table}

\subsection{Input Granularity}
\label{exp:claim_level_exp}
\paragraph{Claim-level}
Table \ref{tab:wice} shows the performance of different decomposition methods combined with two verifiers (AlignScore and Minicheck) on the WICE dataset. With AlignScore, Wice and FactScore decompositions showed positive impacts, leading to higher BAcc and F1 scores than the baseline, while VeriScore had a negative effect. However, when using Minicheck verifier, all methods resulted in lower BAcc and F1 compared to the baseline. Complete details and results of the claim-level experiments are provided in Table \ref{tab:claim_level_wice} and \ref{tab:claim_level_claimdecomp} in Appendix~\ref{appendix:exp_gpt4omini}.

\begin{table}[t]
    \centering
    \setlength{\abovecaptionskip}{0.2cm}
    \setlength{\belowcaptionskip}{-0.4cm}
    \small
    \begin{tabular}{llccc}
        \toprule
        \textbf{Verifier} & \textbf{Decomp} & \textbf{Claims} & \textbf{BAcc} & \textbf{F1} \\
        \midrule
        \multirow{4}{*}{Minicheck} & Baseline & 1.00 & 56.84 & 48.10 \\
        & VeriScore & 5.05 & \cellcolor{lightblue}58.97 & \cellcolor{lightblue}67.56 \\
        & Wice & 5.41 & \cellcolor{lightblue}58.81 & \cellcolor{lightblue}68.12 \\
        & FactScore & 8.69 & \cellcolor{lightred}56.29 & \cellcolor{lightblue}67.51 \\
        \midrule
        \multirow{4}{*}{GPT-4o-mini} & Baseline & 1.00 & 65.28 & 71.56 \\
        & VeriScore & 5.05 & \cellcolor{lightblue}67.28 & \cellcolor{lightred}69.39 \\
        & Wice & 5.41 & \cellcolor{lightred}64.42 & \cellcolor{lightred}64.52 \\
        & FactScore & 8.69 & \cellcolor{lightred}57.86 & \cellcolor{lightred}54.34 \\
        \bottomrule
    \end{tabular}
    \caption{Performance and average number of sub-claims after decomposition for different methods on FELM.}
    \label{tab:felm}
\end{table}

\paragraph{Response-level}
\label{exp:response_level_exp}
Table \ref{tab:felm} presents the performance of different decomposition methods combined with two verifiers (Minicheck and GPT-4o-mini). Compared to the baselines, two trends emerged: (1) A significant increase in F1 when usingmMinicheck, whereas (2) a notable drop in F1 with GPT-4o-mini’s few-shot classification. The detailed results of the response-level experiments are presented in Table \ref{tab:response_level_felm} and \ref{tab:response_level_bingchat} in Appendix~\ref{appendix:exp_gpt4omini}. 

For Minicheck, the improvement from decomposition is significantly more pronounced in the response-level experiments compared to the claim-level experiments. These results indicate that the effectiveness of decomposition is highly dependent on the input granularity.

\subsection{Strong \& Weak Verifier}
\label{sec:strong_weak}
Based on baseline performances in Table \ref{tab:wice} and \ref{tab:felm}, Minicheck demonstrates stronger verification capabilities compared to AlignScore, yet remains less effective than GPT-4o-mini. Our analysis reveals that decomposition generally benefits weaker verifiers, while it tends to negatively affect stronger verification systems. These findings align with prior studies: \citet{tang-etal-2024-minicheck} observe no improvement with decomposition for Minicheck, whereas \citet{kamoi-etal-2023-wice} report gains using a weaker NLI model. Similarly, \citet{zhao2024felm} find claim-based methods advantageous for ChatGPT but detrimental for GPT-4. Despite these isolated observations, no systematic exploration has been conducted. Our study identifies a strong correlation between verifier strength and the impact of decomposition, with further discussion on the trade-offs provided in Section \ref{sec:trade_off}.

\section{What Errors May Decomposition Introduce?}
\label{sec:decomp_errors}
While previous studies~\citep{wanner-etal-2024-closer, min-etal-2023-factscore, song2024veriscore} have identified `atomicity', `coverage', and `verifiability' as key factors in decomposition, less focus has been placed on errors that may harm downstream fact-checking performance. To explore this, we manually inspect cases where the baseline was correct but decomposition led to incorrect outcomes.

\subsection{Error Types}
\label{sec:error_types}
Upon inspection, we identify the following categories of decomposition errors that can jeopardize downstream fact-checking performance:

\paragraph{A. Omission of Context Information}
Excluding critical elements necessary for accurately understanding the input. This category encompasses: 

\begin{itemize}
    \item \emph{Missing Core Claims or Key Details.} The exclusion of essential background details that provide necessary context for understanding the input. Without these elements, sub-claims can become incomplete or misleading.
    \item \emph{Missing Logical Relationships.} Omission of relationships, such as causal, comparative, or contrastive relationship, which explain how different parts of the input relate to one another. These relationships are crucial for understanding the interactions within the original input text.
\end{itemize}

\paragraph{B. Ambiguity}
Decomposing into sub-claims that are unclear or vague can result in multiple interpretations, hindering accurate verification. This refers to \emph{Vague Language}, the use of terms or references that lack specificity, making the claim unclear. This has been exemplified by~\citet{wei2024long}. Examples include:
\begin{enumerate}
    \item Ambiguous pronouns that lack clear referents (e.g., "his," "they," "her").
    \item Vague references to unspecified entities (e.g., "this event," "the research," "the invention").
    \item Incomplete names (e.g., "Jeff..." or "Bezos..." instead of "Jeff Bezos").
\end{enumerate}

\paragraph{C. Over-Decomposition}
Excessive fragmentation of the claim into redundant sub-claims or repeated information, resulting in increased complexity and potential misinterpretation of the original meaning. This category involves:
\begin{itemize}
    \item \emph{Redundant Information.} Repetition of information that does not provide additional value, including sub-claims that reiterate the same content without offering distinct insights.
    \item \emph{Excessive Fragmentation.} Breaking down the input into too many sub-claims that over-explain basic facts, resulting in high complexity and potential misinterpretation of the original meaning.
\end{itemize}

\paragraph{D. Alteration of Original Meaning}
Introducing excessive, fabricated, or contradictory information that changes the original meaning of the claim. This includes misrepresentation or adding elements not present in the original input.

Examples of each error types are provided in Appendix~\ref{appendix:error_type_examples}.

\begin{figure*}
\centering
\setlength{\abovecaptionskip}{0.2cm}
\setlength{\belowcaptionskip}{-0.4cm}
\includegraphics[width=1.\textwidth]{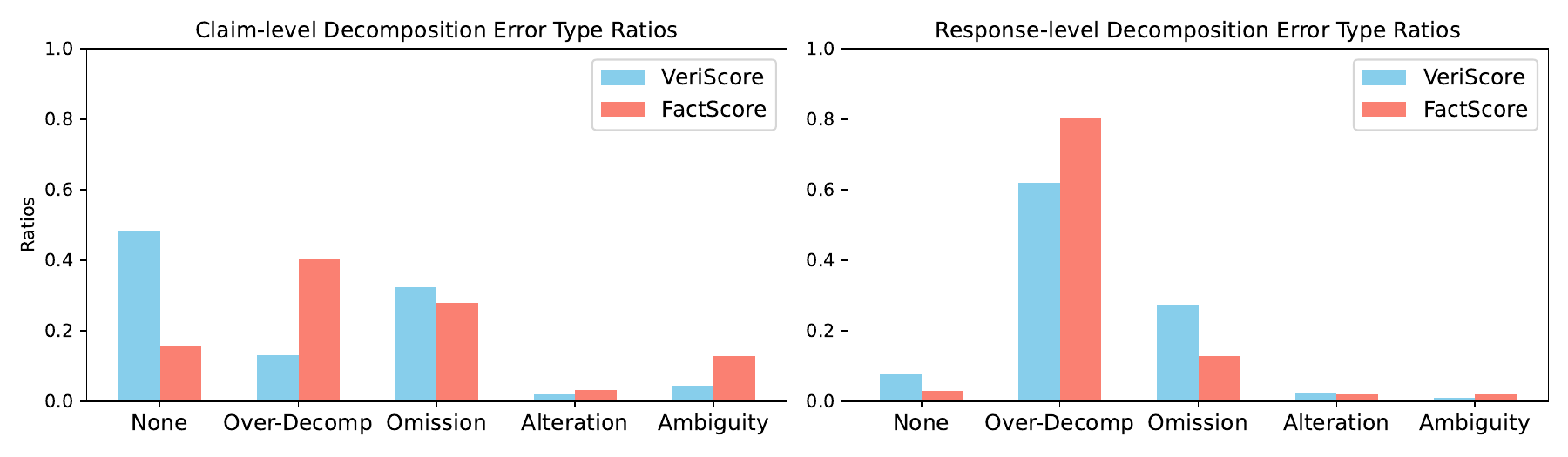}
\caption{Error distribution of using FactScore/VeriScore decomposition on claim-level(WICE, CLAIMDECOMP) and response-level(FELM, BINGCHAT) datasets.}
\label{fig:error_type_dist}
\end{figure*}

\subsection{Error Distribution}
For error detection, given the input text and its decomposed sub-claims, we prompt GPT-4o in a few-shot manner to reason first then determine the errors involved. Details of the few-shot template curation can be found in Appendix~\ref{appendix:prompt_detection}.

Figure~\ref{fig:error_type_dist} shows the error distributions of FactScore and VeriScore decompositions on claim and response-level data. We visualize with these two methods due to their distinct designs: FactScore prioritizes atomicity, breaking input into many atomic sub-claims, while VeriScore focuses on verifiability, resulting in fewer but verifiable sub-claims. The key observations are two-fold:

\paragraph{Method-wise} FactScore exhibits significantly more over-decomposition errors due to excessive fragmentation, leading to high redundancy and potential alteration of the original meaning. In contrast, VeriScore tends to omit contextual information, as its focus on verifiability may overlook important details or relationships in the context. The example in Figure~\ref{fig:pipline} also illustrates the error tendencies specific to each method.

\paragraph{Granularity-wise} Both claim and response-level decompositions exhibit high ratios of over-decomposition errors. At the claim level, particularly for FactScore, further decomposition of an already concise input often increases ambiguity and leads to altered meaning, which explains the higher ratio of ambiguity errors compared to the response level. This suggests there is a balance point where an input is sufficiently verifiable and atomic, and further decomposition leads to altered meaning.

\subsection{Reflection}
Previous studies have shown that prompting models to reflect on and refine their own responses can improve outcomes~\citep{madaan2024self}. Similarly, we tune the error detection prompt to use reasoning and verdict as feedback for refining the initial decomposition. The prompt used is provided in Appendix~\ref{appendix:prompt_reflection}. Table \ref{tab:claimdecomp_bingchat_selfreflect} compares downstream performance using VeriScore and FactScore, with and without reflection, across three datasets. The results show that reflecting on decomposition errors enhances performance,  further supporting the value of the decomposition error categorization.

\begin{table}[]
    \centering
    \setlength{\abovecaptionskip}{0.2cm}
    \setlength{\belowcaptionskip}{-0.4cm}
    \small
    \begin{tabular}{llcc}
        \toprule
        \textbf{Dataset} & \textbf{Decomp} & \textbf{BAcc} & \textbf{F1} \\
        \midrule
        \multirow{4}{*}{CLAIMDECOMP} & VeriScore & 59.37 & 48.03 \\
        & VeriScore$_{\text{reflect}}$ & \textbf{59.91} & \textbf{48.32} \\
        & FactScore & 58.06 & 40.35 \\
        & FactScore$_{\text{reflect}}$ & \textbf{58.56} & \textbf{46.42} \\
        \cmidrule{1-4}
        \multirow{4}{*}{FELM} & VeriScore & 67.28 & 69.39 \\
        & VeriScore$_{\text{reflect}}$ & \textbf{68.30}	& \textbf{72.99} \\
        & FactScore & 57.86 & 54.34 \\
        & FactScore$_{\text{reflect}}$ & \textbf{66.95} & \textbf{70.59} \\
        \cmidrule{1-4}
        \multirow{4}{*}{BINGCHAT} & VeriScore & 54.20 & 53.81 \\
        & VeriScore$_{\text{reflect}}$ & \textbf{55.20} & \textbf{66.54} \\
        & FactScore & 49.93 & 24.91 \\
        & FactScore$_{\text{reflect}}$ & \textbf{56.72} & \textbf{67.70} \\
        \bottomrule
    \end{tabular}
    \caption{Performance of VeriScore/FactScore decompositions with GPT-4o-mini verifier on CLAIMDECOMP, FELM and BINGCHAT, with and without reflection.}
    \label{tab:claimdecomp_bingchat_selfreflect}
\end{table}

\section{What Explains the Variability in Fact-Checking Performance?}
\label{sec:trade_off}
While factors contributing to variability and decomposition errors have been identified, the reasons for decomposition-induced performance variability remain unclear. We propose that this variability results from a decomposition trade-off and conduct experiments to empirically investigate it.

\subsection{Explanation}

Let $A(k)$ represent the verifier's accuracy for an input with complexity $k$, we simply use sequence length and the number of claims as a proxy for input complexity\footnote{We assume that the input complexity correlates with sentence length and the number of claims.}. Generally, verification accuracy is expected to decrease as input complexity increases and decomposition breaks down the input into sub-claims with an average lower complexity, denoted $k_d$. The ideal accuracy gain ($\Delta A$) brought by reduced complexity is defined as the difference between evaluating sub-claims with complexity $k_d$ and the original claim with complexity $k_o$: $\Delta A = A(k_d) - A(k_o)$.

When verifying a single claim, retrieval noise (e.g., irrelevant or distracting content) can affect the final performance~\citep{retrnoise}. We assume each retrieval has an associated probability of error (noise) denoted by $e_r$. Decomposing the input into multiple sub-claims and aggregating them leads to cumulative retrieval noise, denoted by $E_r$. Additionally, the decomposition process itself can introduce errors, denoted by $E_d$.

The baseline performance without decomposition can be expressed as:
 \[
A_{\text{baseline}} = A(k_o) \times (1 - e_r),
\]
and the performance with decomposition is:
\begin{align*}
    A_{\text{decomposed}} &= A(k_d) \times (1 - E_d) \times (1 - E_r) \\
                          &= A(k_d) \times (1 - E).
\end{align*}
 
The performance improvement achieved by decomposition in the presence of error can be expressed as follows:
\begin{align*}
\Delta A_{\rm err} &= A_{\text{decomposed}} - A_{\text{baseline}} \\
         &= [A(k_d) \times (1 - E)] - [A(k_o) \times (1 - e_r)].
\end{align*}

This formulation highlights a trade-off within the \textit{Decompose-Then-Verify} pipeline. Specifically, it balances the accuracy gain resulting from the verification of simpler sub-claims (as $A(k_d)$ increases) against the noise introduced by both decomposition and retrieval processes (as $E$ increases).

To answer the variation between `strong' and `weak' verifiers discussed in Section~\ref{sec:strong_weak}, we assume a `strong' verifier as one has higher accuracy while being more robust, with a lower accuracy decay rate as claim complexity increases.
For weak verifiers, decomposition improves performance ($\Delta A_{\rm err} > 0$) when the significant accuracy gain ($\Delta A$) from simpler sub-claims outweighs the increased noise ($E$). 
For strong verifiers, decomposition can degrade performance ($\Delta A_{\rm err} < 0$) if the marginal accuracy gain ($\Delta A$) brought by reduced complexity does not compensate for the increased noise ($E$).

Besides the verifier strength, the accuracy gain can also be affected by input complexity. While keeping the verifier unchanged, a higher complexity for the original input claim should intuitively lead to a larger accuracy gain introduced by decomposition. Next, we seek to empirically verify the influence of the input complexity on the verification accuracy through the following three experiments.

\begin{table}[t]
    \centering
    \setlength{\abovecaptionskip}{0.2cm}
    \setlength{\belowcaptionskip}{-0.4cm}
    \small
    \begin{tabular}{llccc}
        \toprule
        \textbf{Exp} & \textbf{Dataset} & \textbf{Decomp} & \textbf{BAcc} & \textbf{F1} \\
        \midrule
        \multirow{8}{*}{\textbf{Up}} 
        & \multirow{4}{*}{WICE} & Baseline & 80.01 & 72.32 \\
        & & VeriScore & 74.74 & 65.16 \\
        & & Wice & 76.81 & 68.22 \\
        & & FactScore & 71.11 & 59.90 \\
        \cmidrule{2-5}
        & \multirow{4}{*}{WICE$_{\text{long}}$} & Baseline & 55.52 & 24.64 \\
        & & VeriScore & 55.63 & 26.21 \\
        & & Wice & 56.53 & 27.14 \\
        & & FactScore & 55.88 & 25.71 \\
        \midrule
        \multirow{8}{*}{\textbf{Down}} 
        & \multirow{4}{*}{FELM} & Baseline & 56.84 & 48.10 \\
        & & VeriScore & 58.97 & 64.45 \\
        & & Wice & 58.81 & 68.12 \\
        & & FactScore & 56.29 & 67.51 \\
        \cmidrule{2-5}
        & \multirow{4}{*}{FELM$_{\text{short}}$} & Baseline & 62.88 & 78.09 \\
        & & VeriScore & 61.56 & 78.75 \\
        & & Wice & 61.87 & 79.39 \\
        & & FactScore & 59.81 & 78.87 \\
        \bottomrule
    \end{tabular}
    \caption{Complexity scaling up/down experiments.}
    \label{tab:combined_up_down}
\end{table}

\subsection{Complexity Scale Up} In Section~\ref{exp:claim_level_exp}, we observe that decomposition leads to a performance decline on WICE with Minicheck, this decline is likely due to the marginal accuracy gain ($\Delta A$) from decomposition being insufficient to offset the increased noise. To explore whether a larger $\Delta A$ can mitigate the noise and improve performance, we increase the input complexity.
For data processing, each WICE annotation includes a claim and its preceding sentences in the section. We concatenate each claim with its preceding sentences, keeping the label unchanged. This scaled-up dataset is referred to as WICE$_{\text{long}}$.

\paragraph{Result}
Table~\ref{tab:combined_up_down} presents the performance on WICE and WICE$_{\text{long}}$. On WICE$_{\text{long}}$, decomposition methods show improved performance compared to the baseline, supporting our hypothesis that decomposition becomes more beneficial as input complexity increases. Specifically, the results indicate that the accuracy gain ($\Delta A$) from breaking down complex inputs outweighs the additional noise introduced, leading to improved overall performance when handling more complex input.

\subsection{Complexity Scale Down} As shown in Section~\ref{exp:response_level_exp}, decomposition greatly improved performance on FELM with Minicheck, likely because the substantial accuracy gain ($\Delta A$) outweighed the increased noise. Thus, we scale down input complexity and test if this leads to any performance degradation.
Each response in FELM includes fine-grained annotations, such as labels and comments. We segment these into a shorter claim-level dataset, referred to as FELM$_{\text{short}}$.

\paragraph{Result}
Table~\ref{tab:combined_up_down} also presents the performance on FELM and FELM$_{\text{short}}$. Unlike the improvements seen on FELM, the performance gain from decomposition diminishes on FELM$_{\text{short}}$, with a degradation in BAcc. This suggests that as input complexity decreases, the accuracy gain from decomposition may not offset the increased noise, leading to reduced performance.

\begin{figure}[t]
\centering
\setlength{\abovecaptionskip}{0.2cm}
\setlength{\belowcaptionskip}{-0.4cm}
\includegraphics[width=0.5\textwidth]{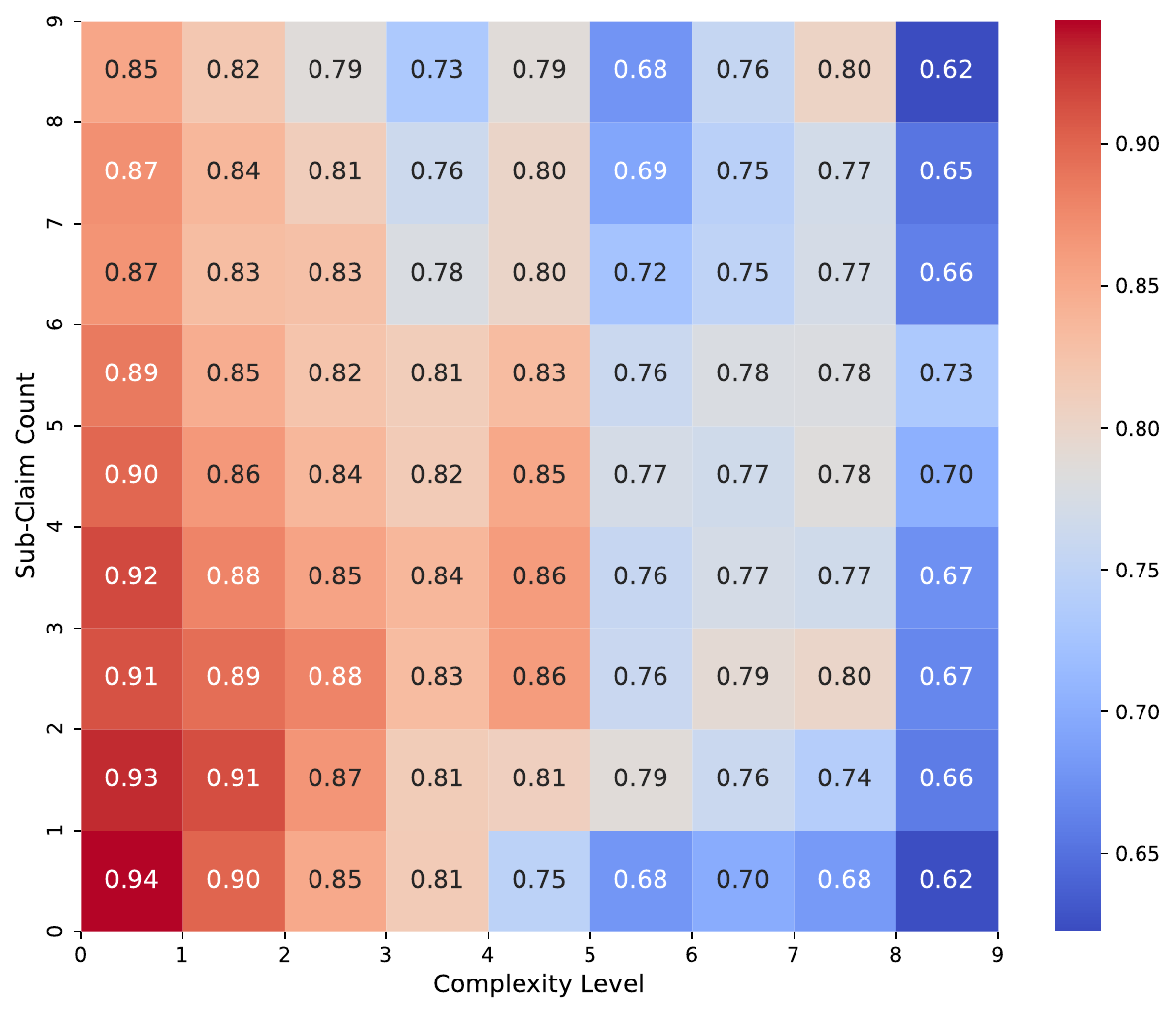}
\caption{The heatmap illustrates F1-scores across varying decomposed sub-claim counts and input complexity levels. Performance generally improves initially as the number of decomposed sub-claims increases, followed by a decline. We find that ensuring the number of sub-claims remains below the input complexity level helps sustain the positive effects of decomposition.} 
\label{fig:heatmap_claim_complexity}
\end{figure}

\subsection{Diverse Complexity Scaling}
To investigate the trade-off in a more fine-grained complexity scenario, we use claim annotations from BINGCHAT. Each response-level entry contains multiple single-sentence claim annotations. We generate combinations of these claims while preserving their original order, with the number of claims serving as a proxy for input complexity. For example, if a response has three claims ($C_1$, $C_2$, $C_3$), the resulting combinations are ($C_1$, $C_2$, $C_3$, $C_1C_2$, $C_2C_3$, $C_1C_2C_3$), yielding complexities ranging from 1 to 3. We sampled 3,000 combinations, with complexities varying from 1 to 9.

The increased noise ($E$) includes decomposition noise ($E_d$) and the accumulation of retrieval noise ($E_r$), which is naturally related to the number of sub-claims ($N$). To control $N$, we instructe GPT-4o to decompose each input into exactly $N$ sub-claims. We test with $N$ ranging from 2 to 9, with 1 indicating no decomposition. The prompt used is in Appendix \ref{appendix:prompt_spec_n}.

\paragraph{Result}
Figure \ref{fig:heatmap_claim_complexity} shows the heatmap of F1 for different sub-claims number with respect to different complexity levels. The heatmap empirically show the existence of the decomposition trade-off. At complexity level 1, the performance degrade as the number of sub-claims increase. For higher complexity levels, as sub-claims increase, the impact of decomposition gradually shift from positive to negative. This suggests that while the initial accuracy gains from decomposition outweigh the noise, further increases in sub-claims lead to diminishing returns as noise surpasses the gains. Additionally, for each level, the maximum F1 is observed when the number of the decomposed sub-claim is less than or equal to the complexity level. Since we correlate input sentence count with complexity in data curation, it suggests that limiting the number of the decomposed sub-claims to be lower than the number of input sentences for higher-complexity input could be helpful to maintain a positive decomposition effect.

\subsection{Summary}
The complexity scaling experiments reveal the inherent trade-off in the \textit{Decompose-Then-Verify} pipeline. While decomposition shows beneficial in handling complex inputs, where accuracy gains outweigh the noise introduced, it becomes less effective for simpler inputs, where the additional noise negates the benefits. Moreover, as sub-claims number increases, the impact of decomposition could shift from positive to negative. The findings highlight the need to balance decomposition granularity. To maximize its effectiveness, more efforts should focus on improving decomposition strategies to better handle varying input complexities and minimize errors introduced in the process.

\section{Conclusion}

We present a comprehensive analysis of decomposition's impact on downstream performance in the \textit{Decompose-Then-Verify} pipeline. We explicitly show the fact-checking performance variations across decomposition methods, input granularities, and verifiers. Through detailed inspection,  we categorize decomposition errors and reveal their distribution across methods and granularities. Furthermore, we reveal a trade-off between the accuracy gains from decomposition and the noise introduced by decomposition, offering an explanation for the inconsistencies reported in prior studies.

\section{Limitations}
This analysis focuses on a binary fact-checking classification task and does not address multi-class classification tasks with additional veracity labels, such as `Partially Supported'. We did not explore pipelines that transform decomposed claims into question-like queries for retrieving answers from a knowledge source, instead of using quantified entailment scores as done here. We plan to explore these aspects in future research. 

In our trade-off analysis, retrieval noise ($E_r$) and decomposition noise ($E_d$) were not explicitly quantified due to the complexity of accounting for various factors. In practice, decomposition could result in sub-claims with varying complexities. To simplify the analysis, we assumed an average sub-claim complexity of $k_i$, which is generally lower than the original input complexity $k_c$. This assumption is a relaxed approximation for the sake of clarity. We empirically demonstrated and visualized the existence of the trade-off. Additionally, in the complexity scaling experiments, we used input length as a proxy for complexity, though other methods for representing input complexity could be explored.


\bibliography{custom}

\newpage
\appendix
\label{sec:appendix}

\onecolumn
\section{Experimental Setting}

\subsection{Dataset Statistics}
\begin{table}[h]
    \centering
    \begin{tabular}{lccc}
        \toprule
        \textbf{Dataset} & \textbf{Splits} & \textbf{Samples} & \textbf{Tokens}\\
        \midrule
        WICE & test & 382 & 33.65  \\
        CLAIMDECOMP & dev/test & 400 & 35.82  \\
        FELM & WK & 184 & 66.74  \\
        BINGCHAT & all & 396 & 351.9  \\
        \bottomrule
    \end{tabular}
    \caption{Dataset statistics.}
    \label{tab:dataset_statistics}
\end{table}

\subsection{Dataset Preprocessing}
\label{appendix:dataset_proc}
\subsubsection{WICE}
We used the test split of WICE dataset along with its provided Wikipedia corpus. The original label space of WICE is \{`SUPPORTED', `PARTIALLY-SUPPORTED', `NOT-SUPPORTED'\}. As our studied task focus on binary classification, we regard `PARTIALLY-SUPPORTED' and `NOT-SUPPORTED' as `Unsupported', and `SUPPORTED' as `Supported' in our case. 

\subsubsection{CLAIMDECOMP}
We used both the test and dev splits of CLAIMDECOMP dataset. The original label space of CLAIMDECOMP is \{`pants-on-fire', `false', `barely-true', `half-true', `mostly-true', `true'\}. We regard \{`pants-on-fire', `false', `barely-true', `half-true'\} as `Unsupported', and \{`mostly-true', `true'\} as `Supported'.

\subsubsection{FELM}
We used the `World Knowledge'(WK) subset of FELM dataset. The subset contains ChatGPT's responses for questions related with history, society, common
sense and so on. We employed decontextualization by instructing GPT-4o with the prompt shown in Appendix~\ref{appendix:prompt_decontext}.

\subsubsection{BINGCHAT}
We used the whole BINGCHAT dataset and employed the same decontextualization with the prompt shown in Appendix~\ref{appendix:prompt_decontext}. For label processing, For label processing, if any claim within a response is annotated as `refuted', we regard this response label as `Unsupported', otherwise `Supported'.

\subsubsection{WICE$_{\text{long}}$} In the WICE dataset, each claim is accompanied by its preceding sentences from the same Wikipedia section. We concatenated these preceding sentences with the claim to form a longer input, while keeping the original label unchanged.

\subsubsection{FELM$_{\text{short}}$} In the FELM dataset, each response contains fine-grained annotations, including segment labels, annotator comments, and reference links. We prompted GPT-4o to decompose each segment into two claims using the prompt in Appendix~\ref{appendix:prompt_spec_n}. The label for each claim was then assigned based on the corresponding segment label and annotator comments.

\subsubsection{BINGCHAT$_{\text{diverse}}$}
Each entry in BINGCHAT contains multiple claim annotations. We generated combinations of claims without altering their order, using the number of claims to represent input complexity. We sampled 3,000 combinations, varying complexity from 1 to 9. For label processing, if any claim within a combination is `refuted', we regard this combination as `Unsupported', otherwise `Supported'.

\subsubsection{Decontextualization}
Decontextualization~\citep{choi-etal-2021-decontextualization, gunjal2024molecular} refers to the process that given a text together with its context and rewriting it to be interpretable out of context. This step is crucial for processing response-level data, ensuring that responses are understandable without relying on the original question. We perform decontextualization by prompting GPT-4o~\citep{gpt4oreport}. The prompt used is provided in Appendix~\ref{appendix:prompt_decontext}.

\subsection{Pipeline Settings}
\label{appendix:verifier_details}
\subsubsection{AlignScore \& Minicheck}
Both AlignScore~\citep{zha-etal-2023-alignscore} and Minicheck~\citep{tang-etal-2024-minicheck} provide models and official script for calculating entailment score ranging from 0 to 1. Specifically, we used AlignScore-large and Bespoke-MiniCheck-7B for our experiments. Following~\citet{kamoi-etal-2023-wice}, for each sub-claim and its retrieved evidences, we used the maximum score to combine the entailment scores and regard it as the entailment score for the sub-claim.

\subsubsection{Retrieval Module}  
For CLAIMDECOMP, FELM, and BINGCHAT, we used Google Search\footnote{\url{https://serper.dev/}} to retrieve the top 10 results as evidence, following prior practices~\citep{chern2023factool, song2024veriscore}. For WICE, we used an embedding model to convert Wikipedia articles into vector representations, stored in a Chroma vector database\footnote{\url{https://github.com/chroma-core/chroma}} for retrieval. The embedding model used was `dunzhang/stella\_en\_400M\_v5', a model further trained on gte-large-en-v1.5~\citep{zhang2024mgte}.

\subsubsection{GPT-4o-mini as Verifier}
We utilize the verification module from \citet{song2024veriscore}, which provides a few-shot classification prompt to determine whether an input text is `Supported' or `Unsupported' based on a list of Google search results. The entailment score is computed by cleaning the model's response and checking for target tokens such as "supported" or "unsupported" in the log probabilities. If the target token is found, its log probability is used; otherwise, the cumulative log probability of all tokens is applied. The final score is the exponentiated probability: for `Supported', it is the probability itself, and for `Unsupported', it is $1 - \text{probability}$.

\subsection{Hardware \& Library}
For experiments using AlignScore or Minicheck as the verifier, we use an A100 40GB GPU. AlignScore is hosted locally using NVIDIA NeMo Guardrails\footnote{\url{https://docs.nvidia.com/nemo-guardrails/}}, while for Minicheck, we utilize the official Python package\footnote{\url{https://github.com/Liyan06/MiniCheck}} from the official repository. Additionally, we conduct experiments with Llama-3.1-8B\footnote{meta-llama/Meta-Llama-3.1-8B} as the backbone for the decomposer, hosting the model with vllm~\citep{kwon2023efficient}.

\subsection{Inference Parameter Settings}
For all operations involving LLM inference, we explicitly set the temperature to 0, while keeping all other parameters at their default values.

\newpage
\onecolumn
\section{Experimental Results}
\subsection{Default: GPT-4o-mini as Decomposer}
\label{appendix:exp_gpt4omini}
\begin{table*}[!htbp]
    \centering
    \small
    \begin{tabular}{llccccc}
        \toprule
        \textbf{Verifier} & \textbf{Decomp} & \textbf{Claims} & \textbf{BAcc} & \textbf{F1} & \textbf{Precision} & \textbf{Recall} \\
        \midrule
        \multirow{4}{*}{AlignScore} 
        & Baseline & 1.00 & 54.80 & 40.78 & 36.11 & 46.85\\
        & VeriScore & 2.63 & 52.17 & 38.81 & 33.12 & 46.85 \\
        & Wice & 3.10 & 56.26 & 42.23 & 38.17 & 45.05\\
        & FactScore & 3.91 & 56.87 & 42.74 & 38.68 & 47.75\\
        \midrule
        \multirow{4}{*}{Minicheck} 
        & Baseline & 1.00 & 80.01 & 72.32 & 71.68 & 72.97 \\
        & VeriScore & 2.62 & 74.74 & 65.16 & 65.45 & 64.86 \\
        & Wice & 3.10 & 76.81 & 68.22 & 70.87 & 65.77 \\
        & FactScore & 3.91 & 71.11 & 59.90 & 68.60 & 53.15 \\
        \bottomrule
    \end{tabular}
    \caption{Complete pipeline performance (BAcc/F1/Precision/Recall) and average number of sub-claims for different methods(GPT-4o-mini as decomposer) on the WICE dataset. GPT-4o-mini as verifier is excluded because its few-shot prompt from VeriScore~\citep{song2024veriscore} is specifically designed for Google Search results, while the WICE dataset uses Wikipedia article retrieval.}
    \label{tab:claim_level_wice}
\end{table*}

\begin{table*}[!htbp]
    \centering
    \small
    \begin{tabular}{llccccc}
        \toprule
        \textbf{Verifier} & \textbf{Decomp} & \textbf{Claims} & \textbf{BAcc} & \textbf{F1} & \textbf{Precision} & \textbf{Recall} \\
        \midrule
        \multirow{4}{*}{AlignScore} 
        & Baseline & 1.00 & 51.38 & 44.09 & 33.47 & 64.62\\
        & VeriScore & 1.96 & 53.93 & 45.25 & 35.53 & 62.31\\
        & Wice & 2.86 & 55.17 & 42.81 & 37.87 & 49.23\\
        & FactScore & 4.02 & 56.52 & 44.74 & 39.08	& 52.31\\
        \midrule
        \multirow{4}{*}{Minicheck} 
        & Baseline & 1.00 & 59.50 & 47.42 & 42.86 & 53.08 \\
        & VeriScore & 1.96 & 59.34 & 47.65 & 42.26 & 54.62\\
        & Wice & 2.86 & 59.22 & 48.23 & 41.44 & 57.69\\
        & FactScore & 4.02 & 56.72 & 45.10 & 39.20 & 53.08\\
        \midrule
        \multirow{4}{*}{GPT-4o-mini} 
        & Baseline & 1.00 & 58.43 & 49.13 & 39.35 & 65.38 \\
        & VeriScore & 1.96 & 59.37 & 48.03 & 41.95 & 56.15\\
        & Wice & 2.86 & 58.19 & 43.14 & 44.00 & 42.31\\
        & FactScore & 4.02 & 58.06 & 40.35 & 46.94 & 35.38\\
        \bottomrule
    \end{tabular}
    \caption{Complete pipeline performance (BAcc/F1/Precision/Recall) and average number of sub-claims for different methods (GPT-4o-mini as decomposer) on the CLAIMDECOMP dataset.}
    \label{tab:claim_level_claimdecomp}
\end{table*}

\begin{table*}[!htbp]
    \centering
    \small
    \begin{tabular}{llccccc}
        \toprule
        \textbf{Verifier} & \textbf{Decomp} & \textbf{Claims} & \textbf{BAcc} & \textbf{F1} & \textbf{Precision} & \textbf{Recall} \\
        \midrule
        \multirow{4}{*}{AlignScore} & Baseline & 1.00 & 50.87 & 45.88 & 54.93 & 39.39 \\
        & VeriScore & 5.05 & 56.11 & 63.26 & 58.62 & 68.69 \\
        & Wice & 5.41 & 50.98 & 60.00 & 54.55 & 66.67 \\
        & FactScore & 8.69 & 54.52 & 64.32 & 57.03 & 73.74 \\
        \midrule
        \multirow{4}{*}{Minicheck} & Baseline & 1.00 & 56.84 & 48.10 & 64.41 & 38.38 \\
        & VeriScore & 5.05 & 58.97 & 67.56 & 60.32 & 76.77 \\
        & Wice & 5.41 & 58.81 & 68.12 & 60.00 & 78.79 \\
        & FactScore & 8.69 & 56.29 & 67.51 & 57.97 & 80.81 \\
        \midrule
        \multirow{4}{*}{GPT-4o-mini} & Baseline & 1.00 & 65.28 & 71.56 & 65.55	& 78.79 \\
        & VeriScore & 5.05 & 67.28 & 69.39 & 70.10 & 68.69 \\
        & Wice & 5.41 & 64.42 & 64.52 & 68.97 & 60.61 \\
        & FactScore & 8.69 & 57.86 & 54.34 & 63.51 & 47.47 \\
        \bottomrule
    \end{tabular}
    \caption{Complete pipeline performance (BAcc/F1/Precision/Recall) and average number of sub-claims for different methods (GPT-4o-mini as decomposer) on the FELM dataset.}
    \label{tab:response_level_felm}
\end{table*}

\begin{table*}[!htbp]
    \centering
    \small
    \begin{tabular}{llccccc}
        \toprule
        \textbf{Verifier} & \textbf{Decomp} & \textbf{Claims} & \textbf{BAcc} & \textbf{F1} & \textbf{Precision} & \textbf{Recall} \\
        \midrule
        \multirow{4}{*}{AlignScore} & Baseline & 1.00 & 50.24 & 00.96	& 100.0 & 00.48 \\
        & VeriScore & 23.21 & 51.76 & 47.41 & 54.72 & 41.83 \\
        & Wice & 18.31 & 54.57 & 34.97 & 64.10 & 24.04\\
        & FactScore & 57.64 & 53.04 & 48.63 & 56.33 & 42.79\\
        \midrule
        \multirow{4}{*}{Minicheck} & Baseline & 1.00 & 52.23 & 32.99 & 57.83 & 23.08 \\
        & VeriScore & 23.21 & 56.94 & 64.99 & 57.62 & 74.52 \\
        & Wice & 18.31 & 50.48 & 42.98 & 53.19 & 36.06\\
        & FactScore & 57.64 & 50.69 & 65.82 & 52.92 & 87.02\\
        \midrule
        \multirow{4}{*}{GPT-4o-mini} & Baseline & 1.00 & 54.59 & 66.67 & 55.41 & 83.65 \\
        & VeriScore & 23.21 & 54.20 & 53.81 & 56.99 & 50.96 \\
        & Wice & 18.31 & 55.91 & 60.59 & 57.58 & 63.94\\
        & FactScore & 57.64 & 49.64 & 25.36 & 51.47 & 16.83\\
        \bottomrule
    \end{tabular}
    \caption{Complete pipeline performance (BAcc/F1/Precision/Recall) and average number of sub-claims for different methods (GPT-4o-mini as decomposer) on the BINGCHAT dataset.}
    \label{tab:response_level_bingchat}
\end{table*}

\clearpage
\subsection{Llama-3.1-8B as Decomposer}
 We also include results using Llama-3.1-8B as the decomposer. Overall, similar trends were observed, as shown in Appendix~\ref{appendix:exp_gpt4omini}. Interestingly, on certain datasets like CLAIMDECOMP, Llama-3.1-8B performed on par with, or even outperformed, GPT-4o-mini. This suggests the potential of improving open-source models to surpass proprietary models in claim decomposition, and possibly in full fact-checking pipelines.

\begin{table*}[!htbp]
    \centering
    \small
    \begin{tabular}{llccccc}
        \toprule
        \textbf{Verifier} & \textbf{Decomp} & \textbf{Claims} & \textbf{BAcc} & \textbf{F1} & \textbf{Precision} & \textbf{Recall} \\
        \midrule
        \multirow{4}{*}{AlignScore} 
        & Baseline & 1.00 & 54.80 & 40.78 & 36.11 & 46.85\\
        & VeriScore & 2.42 & 51.81 & 38.83 & 32.72 & 47.75 \\
        & Wice & 2.92 & 55.22 & 40.00 & 37.21 & 43.24\\
        & FactScore & 3.76 & 52.44 & 38.13 & 33.56 & 44.14\\
        \midrule
        \multirow{4}{*}{Minicheck} 
        & Baseline & 1.00 & 80.01 & 72.32 & 71.68 & 72.97 \\
        & VeriScore & 2.42 & 74.00 & 64.19 & 66.35 & 62.16 \\
        & Wice & 2.92 & 74.69 & 65.14 & 66.36 & 63.96 \\
        & FactScore & 3.76 & 70.89 & 59.72 & 63.00 & 56.76\\
        \bottomrule
    \end{tabular}
    \caption{Complete pipeline performance (BAcc/F1/Precision/Recall) and average number of sub-claims for different methods(Llama-3.1-8B as decomposer) on the WICE dataset.}
\end{table*}

\begin{table*}[!htbp]
    \centering
    \small
    \begin{tabular}{llccccc}
        \toprule
        \textbf{Verifier} & \textbf{Decomp} & \textbf{Claims} & \textbf{BAcc} & \textbf{F1} & \textbf{Precision} & \textbf{Recall} \\
        \midrule
        \multirow{4}{*}{AlignScore} 
        & Baseline & 1.00 & 51.38 & 44.09 & 33.47 & 64.62\\
        & VeriScore & 1.67 & 53.01 & 44.63 & 34.76 & 62.31\\
        & Wice & 2.75 & 52.75 & 40.78 & 35.20 & 48.46\\
        & FactScore & 5.18 & 57.18 & 43.97 & 40.79 & 47.69\\
        \midrule
        \multirow{4}{*}{Minicheck} 
        & Baseline & 1.00 & 59.50 & 47.42 & 42.86 & 53.08 \\
        & VeriScore & 1.67 & 58.60 & 47.02 & 41.28 & 54.62\\
        & Wice & 2.75 & 59.47 & 47.02 & 43.23 & 51.54\\
        & FactScore & 5.18 & 61.52 & 49.10 & 46.26 & 52.31\\
        \midrule
        \multirow{4}{*}{GPT-4o-mini} 
        & Baseline & 1.00 & 58.43 & 49.13 & 39.35 & 65.38 \\
        & VeriScore & 1.67 & 61.94 & 51.85 & 43.30 & 64.62\\
        & Wice & 2.75 & 58.59 & 44.02 & 44.19 & 43.85\\
        & FactScore & 5.18 & 59.57 & 42.48 & 50.00 & 36.92\\
        \bottomrule
    \end{tabular}
    \caption{Complete pipeline performance (BAcc/F1/Precision/Recall) and average number of sub-claims for different methods (Llama-3.1-8B as decomposer) on the CLAIMDECOMP dataset.}
\end{table*}

\begin{table*}[!htbp]
    \centering
    \small
    \begin{tabular}{llccccc}
        \toprule
        \textbf{Verifier} & \textbf{Decomp} & \textbf{Claims} & \textbf{BAcc} & \textbf{F1} & \textbf{Precision} & \textbf{Recall} \\
        \midrule
        \multirow{4}{*}{AlignScore} & Baseline & 1.00 & 50.87 & 45.88 & 54.93 & 39.39 \\
        & VeriScore & 6.65 & 49.97 & 58.72 & 53.78 & 64.65 \\
        & Wice & 4.54 & 49.80 & 56.46 & 53.64 & 59.60 \\
        & FactScore & 8.90 & 50.56 & 58.99 & 54.24 & 64.65 \\
        \midrule
        \multirow{4}{*}{Minicheck} & Baseline & 1.00 & 56.84 & 48.10 & 64.41 & 38.38 \\
        & VeriScore & 6.65 & 59.56 & 67.86 & 60.80 & 76.77 \\
        & Wice & 4.54 & 56.53 & 64.22 & 58.82 & 70.71 \\
        & FactScore & 8.90 & 53.93 & 64.04 & 56.59 & 73.74 \\
        \midrule
        \multirow{4}{*}{GPT-4o-mini} & Baseline & 1.00 & 65.28 & 71.56 & 65.55	& 78.79 \\
        & VeriScore & 6.65 & 62.49 & 64.25 & 65.96 & 62.63 \\
        & Wice & 4.54 & 62.15 & 62.77 & 66.29 & 59.60 \\
        & FactScore & 8.90 & 57.43 & 52.94 & 63.38 & 45.45 \\
        \bottomrule
    \end{tabular}
    \caption{Complete pipeline performance (BAcc/F1/Precision/Recall) and average number of sub-claims for different methods (Llama-3.1-8B as decomposer) on the FELM dataset.}
\end{table*}

\begin{table*}[!htbp]
    \centering
    \small
    \begin{tabular}{llccccc}
        \toprule
        \textbf{Verifier} & \textbf{Decomp} & \textbf{Claims} & \textbf{BAcc} & \textbf{F1} & \textbf{Precision} & \textbf{Recall} \\
        \midrule
        \multirow{4}{*}{AlignScore} & Baseline & 1.00 & 50.24 & 00.96	& 100.0 & 00.48 \\
        & VeriScore & 28.00 & 53.29 & 42.20 & 57.98 & 33.17\\
        & Wice & 11.59 & 48.84 & 25.09 & 49.30 & 16.83\\
        & FactScore & 72.56 & 51.97 & 37.30 & 56.31 & 27.88\\
        \midrule
        \multirow{4}{*}{Minicheck} & Baseline & 1.00 & 52.23 & 32.99 & 57.83 & 23.08 \\
        & VeriScore & 28.00 & 52.15 & 52.02 & 54.79 & 49.52\\
        & Wice & 11.59 & 50.78 & 38.77 & 53.85 & 30.29\\
        & FactScore & 72.56 & 52.82 & 58.24 & 54.89 & 62.02\\
        \midrule
        \multirow{4}{*}{GPT-4o-mini} & Baseline & 1.00 & 54.59 & 66.67 & 55.41 & 83.65 \\
        & VeriScore & 28.00 & 53.95 & 50.80 & 57.23 & 45.67\\
        & Wice & 11.59 & 55.12 & 62.04 & 56.52 & 68.75\\
        & FactScore & 72.56 & 50.56 & 22.81 & 54.55 & 14.42\\
        \bottomrule
    \end{tabular}
    \caption{Complete pipeline performance (BAcc/F1/Precision/Recall) and average number of sub-claims for different methods (Llama-3.1-8B as decomposer) on the BINGCHAT dataset.}
\end{table*}

\onecolumn
\section{Decomposition Error Examples}
\label{appendix:error_type_examples}
\begin{table*}[!htbp]
\label{tab:error_examples}
\small
\begin{tabularx}{\textwidth}{|l|X|}
\hline
\textbf{Error Type}& \textbf{Example (Input text \& Error Decomposition)}\\
\hline
Omission (Logical Relationships) & 
\textbf{\#Input:} 

\textit{Due to Imran Khan's criticism of Macron's comments on Islam, French authorities cancelled the visas of 183 Pakistani citizens and deported 118 from the country.}

\textbf{\#Decomposition: } 

- French authorities cancelled the visas of 183 Pakistani citizens.

- French authorities deported 118 Pakistani citizens from the country.

\color{red}{\textbf{\# Justification}

Missing causal relationship that links Imran Khan's criticism of Macron's comments on Islam to the actions of visa cancellation and deportation.} \\
\hline
Omission (Core Claims) & 
\textbf{\#Input:} 

\textit{Other Countries Have Proved that the U.S. FDA Was Wrong to Withdraw Approval Of Hydroxychloroquine as COVID Drug.}

\textbf{\#Decomposition: }

- Other countries have proved something.

- The U.S. FDA withdrew approval of hydroxychloroquine.

- Hydroxychloroquine was considered a COVID drug.

\color{red}{\textbf{\# Justification}

Missing the core claim core claim that other countries proved the U.S. FDA's decision on hydroxychloroquine was incorrect, highlighting a contradiction. } \\

\hline
Ambiguity (Vague Language) & 
\textbf{\#Input:} 

\textit{The bacterium responsible for causing tuberculosis was discovered by German physician and microbiologist Robert Koch in 1882.}

\textbf{\#Decomposition: }

- The bacterium causes tuberculosis.

- The bacterium was discovered by Robert Koch.

- Robert Koch was a German physician.

- Robert Koch was a microbiologist.

- The bacterium was discovered in 1882.

\color{red}{\textbf{\# Justification}

Several sub-claims are not self-contained and are ambiguous. The phrases "The bacterium" and "The discovery" lack clear references when taken individually.} \\

\hline
Over-Decomposition & 
\textbf{\#Input:}  

\textit{As an AI language model, I cannot make subjective statements or predictions about the success or failure of a particular aircraft. However, it is worth noting that the Sukhoi Su-57 has faced some challenges in its development, including delays and budget constraints. Additionally, there are concerns about the aircraft's engine and stealth capabilities. Ultimately, the success of the Su-57 will depend on a variety of factors, including its performance in combat situations and its ability to meet the needs of the Russian military.}

\textbf{\#Decomposition: }

- I am an AI language model.

- I cannot make subjective statements.

- I cannot make predictions about the success of a particular aircraft.

- I cannot make predictions about the failure of a particular aircraft.

- The Sukhoi Su-57 has faced challenges in its development.

- The challenges faced by the Sukhoi Su-57 include delays.

- The challenges faced by the Sukhoi Su-57 include budget constraints.

...

\color{red}{\textbf{\# Justification}

The decomposition should focus on verifiable information about the Sukhoi Su-57's challenges and success factors, excluding redundant statements like "I am an AI language model." Listing too many sub-claims, such as delays and budget constraints separately, results in over-decomposition.} \\

\hline
Alteration of Original Meaning & 
\textbf{\#Input:} 

\textit{President Ferdinand Marcos and Dr. Jose Rizal established and founded the WORLD BANK and International Monetary Fund.}

\textbf{\#Decomposition: }

- President Ferdinand Marcos did not establish or found the WORLD BANK.

- Dr. Jose Rizal did not establish or found the WORLD BANK.

- President Ferdinand Marcos did not establish or found the International Monetary Fund.

- Dr. Jose Rizal did not establish or found the International Monetary Fund.

\color{red}{\textbf{\# Justification}
The original text claims Marcos and Rizal founded the World Bank and IMF. The decomposition incorrectly negates this, altering the text's meaning. } \\

\hline
\end{tabularx}
\end{table*}

\onecolumn
\section{Prompt}

\label{appendix:prompt}
\subsection{Decomposition Error Detection}
\label{appendix:prompt_detection}
As outlined in Section~\ref{sec:decomp_errors}, we manually categorized decomposition error types. Using these definitions, we prompted the OpenAI-o1 model~\citep{o1report} to analyze decomposition errors. We refined its analysis and used it for few-shot demonstrations to enhance error detection.  Additionally, we conducted a small-scale human evaluation by sampling 100 decomposition outcomes from all datasets using FactScore/VeriScore. The evaluation revealed an 83\% agreement rate with the GPT-4o predictions. Below is the format of the detection prompt:

\texttt{%
Fact-checking involves accessing the veracity of a given text, which could be a statement, claim, paragraph, or a generated response from a large language model. \\
For more complex fact-checking tasks, the input text is often broken down into a set of manageable, verifiable and self-contained sub-claim, a process called 'Input Decomposition'. Each sub-claim is required to be *self-contained*, meaning it is completely interpretable without the original text and other sub-claims. Each sub-claim is verified independently, and the results are combined to assess the overall veracity of the original input. The decomposition error categories are defined as below: \\
\# Error Categories in Decomposition\\
---\\
\#\#\# Omission of Context Information\\
Failure to include critical elements necessary for accurate understanding or verification of the claim. This category encompasses:\\
\\
\#\#\#\# Missing Core Claims or Key Details\\
- **Definition**: Exclusion of essential background or situational details that provide the necessary context for understanding the claim. Without these elements, the sub-claims become incomplete or misleading.
\\
\#\#\#\# Missing Logical Relationships\\
- **Definition**: Omission of relationships such as cause-and-effect links (Causal Relationship), comparisons (Comparative Relationship), or contrasts that explain how different parts of the claim relate to each other. These relationships are vital for understanding the interactions within the original claim.\\
\{More definitions...\}\\
\\
Your task is to evaluate whether the following input decomposition is 'Acceptable' or 'Problematic'.\\
Please give your final judgment and support it with your justification. If the decomposition is problematic, identify the error(s) involved.\\
\\
Use tripple backticks to enclose the reasoning process, error type, and judgment. Your response MUST follow this format:\\
\\
\#\#\# Reasoning\\
```\\
Provide a step-by-step explanation of your reasoning.\\
```\\
\\
\#\#\# Error Type\\
```\\
Identify the specific type of error, if any, in the decomposition.\\
```\\
\\
\#\#\# Judgment\\
```\\
Conclude whether the decomposition is 'Acceptable' or 'Problematic'.\\
```\\
\\
Here are some examples:\\
\\
\#\#\# Given text
```\\
Due to Imran Khan's criticism of Macron's comments on Islam, French authorities cancelled the visas of 183 Pakistani citizens and deported 118 from the country.\\
```\\
\\
\#\#\# Decomposition
```\\
- French authorities cancelled the visas of 183 Pakistani citizens.\\
- French authorities deported 118 Pakistani citizens from the country.\\
```\\
\#\#\# Reasoning
```\\
The decomposition breaks the given text into two sub-claims about the actions of the French authorities. However, it fails to incorporate the causal relationship that links Imran Khan's criticism of Macron's comments on Islam to the actions of visa cancellation and deportation. This omission makes the decomposition incomplete, as the causal context is crucial for understanding why these actions were taken.
```\\
\\
\#\#\# Error Type
```\\
Omission of Context Information - Missing Logical Relationships
```\\
\\
\#\#\# Judgment
```\\
Problematic
```\\
\\
==========\\
\\
\#\#\# Given text
```\\
The smallest ocean in the world is the Arctic Ocean. It is located in the northernmost part of the Earth and is surrounded by the land masses of North America, Europe, and Asia. The Arctic Ocean covers an area of about 14.05 million square kilometers.\\
```\\
\\
\#\#\# Decomposition
```\\
- The smallest ocean in the world is the Arctic Ocean.\\
- The Arctic Ocean is surrounded by the land masses of North America.\\
- The Arctic Ocean is surrounded by the land masses of Europe.\\
- The Arctic Ocean is surrounded by the land masses of Asia.\\
- The Arctic Ocean covers an area of about 14.05 million square kilometers.\\
```\\
\\
\#\#\# Reasoning
```\\
The decomposition involves breaking down the fact that the Arctic Ocean is surrounded by North America, Europe, and Asia into three distinct sub-claims. This is an example of excessive fragmentation, as it introduces unnecessary complexity and breaks down information that should be kept together for concise understanding. This fragmentation does not add value and makes the decomposition more cumbersome.\\
\\
Moreover, this fragmentation alters the original meaning. The original text implies that the Arctic Ocean is surrounded by all three continents collectively, but decomposing them into separate sub-claims can suggest an incorrect interpretation that each continent surrounds the ocean independently.\\
\\
Additionally, the decomposition omits the core contextual detail that the Arctic Ocean is located in the northernmost part of the Earth. This omission reduces the completeness of the decomposition and makes it more challenging to verify the claims properly.\\
```\\
\\
\#\#\# Error Type
```\\
- Over-Decomposition: Excessive Fragmentation\\
- Over-Decomposition: Redundant Information\\
- Omission of Context Information: Missing Core Claims or Key Details\\
- Alteration of Original Meaning\\
```\\
\\
==========\\
\\
\{More demonstrations...\}\\
}

\subsection{Decomposition Error Reflection}
\label{appendix:prompt_reflection}
\texttt{
Fact-checking involves accessing the veracity ...\\
\{Error definition same as detection prompt.\}\\
Your task is to evaluate whether the following input decomposition is 'Good' or 'Problematic' or 'No need for decomposition'.\\
\\
Please give your final judgment and support it with your justification. If the decomposition is problematic, identify the type of error involved and suggest the set of refined decomposition when necessary. If the given text is already concise and clear enough and does not need further decomposition, you can use the given claim as the `Refined Decomposition` and mark the Judgement as 'No need for decomposition'. \\
\\
Use tripple backticks to enclose the reasoning process, error type, judgment and refined decomposition. Your response MUST follow this format:\\
\#\#\# Reasoning
```\\
Provide a step-by-step explanation of your reasoning.\\
```\\
\\
\#\#\# Error Type
```\\
Identify the specific type of error, if any, in the decomposition.\\
```\\
\\
\#\#\# Judgment
```\\
Conclude whether the decomposition is 'Good' or 'Problematic' or 'No need for decomposition'.\\
```\\
\\
\#\#\# Refined Decomposition \\
```\\
Provide the refined set of sub-claims.\\
```\\
\\
Here are some examples:\\
\\
\#\#\# Given text
```\\
Due to Imran Khan's criticism of Macron's comments on Islam, French authorities cancelled the visas of 183 Pakistani citizens and deported 118 from the country.\\
```\\
\\
\#\#\# Decomposition
```\\
- French authorities cancelled the visas of 183 Pakistani citizens.\\
- French authorities deported 118 Pakistani citizens from the country.\\
```\\
\#\#\# Reasoning
```\\
The decomposition breaks the given text into two sub-claims about the actions of the French authorities. However, it fails to incorporate the causal relationship that links Imran Khan's criticism of Macron's comments on Islam to the actions of visa cancellation and deportation. This omission makes the decomposition incomplete, as the causal context is crucial for understanding why these actions were taken.
```\\
\\
\#\#\# Error Type
```\\
Omission of Context Information - Missing Logical Relationships
```\\
\\
\#\#\# Judgment
```\\
Problematic
```\\
\\
\#\#\# Refined Decomposition
```\\
- Due to Imran Khan's criticism of Macron's comments on Islam, French authorities cancelled the visas of 183 Pakistani citizens.\\
- Due to Imran Khan's criticism of Macron's comments on Islam, French authorities deported 118 Pakistani citizens from the country.\\
```\\
\\
==========\\
\\
\#\#\# Given text
```\\
The smallest ocean in the world is the Arctic Ocean. It is located in the northernmost part of the Earth and is surrounded by the land masses of North America, Europe, and Asia. The Arctic Ocean covers an area of about 14.05 million square kilometers.\\
```\\
\\
\#\#\# Decomposition
```\\
- The smallest ocean in the world is the Arctic Ocean.\\
- The Arctic Ocean is surrounded by the land masses of North America.\\
- The Arctic Ocean is surrounded by the land masses of Europe.\\
- The Arctic Ocean is surrounded by the land masses of Asia.\\
- The Arctic Ocean covers an area of about 14.05 million square kilometers.\\
```\\
\\
\#\#\# Reasoning
```\\
The decomposition involves breaking down the fact that the Arctic Ocean is surrounded by North America, Europe, and Asia into three distinct sub-claims. This is an example of excessive fragmentation, as it introduces unnecessary complexity and breaks down information that should be kept together for concise understanding. This fragmentation does not add value and makes the decomposition more cumbersome.\\
\\
Moreover, this fragmentation alters the original meaning. The original text implies that the Arctic Ocean is surrounded by all three continents collectively, but decomposing them into separate sub-claims can suggest an incorrect interpretation that each continent surrounds the ocean independently.\\
\\
Additionally, the decomposition omits the core contextual detail that the Arctic Ocean is located in the northernmost part of the Earth. This omission reduces the completeness of the decomposition and makes it more challenging to verify the claims properly.\\
```\\
\\
\#\#\# Error Type
```\\
- Over-Decomposition: Excessive Fragmentation\\
- Over-Decomposition: Redundant Information\\
- Omission of Context Information: Missing Core Claims or Key Details\\
- Alteration of Original Meaning\\
```\\
\\
\#\#\# Refined Decomposition
```\\
- The smallest ocean in the world is the Arctic Ocean.\\
- The Arctic Ocean is located in the northernmost part of the Earth.\\
- The Arctic Ocean is surrounded by the land masses of North America, Europe, and Asia.\\
- The Arctic Ocean covers an area of about 14.05 million square kilometers.\\
```\\
==========\\
\{More demonstrations...\}\\
}

\subsection{Specify Number of Sub-claims}
\label{appendix:prompt_spec_n}
\texttt{
Fact-checking involves accessing the veracity of a given text, which could be a statement, claim, paragraph, or a generated response from a large language model.\\
\\
For more complex fact-checking tasks, the input text is often broken down into a set of manageable, verifiable and self-contained claims, a process called 'Input Decomposition'. Each claim is required to be *self-contained*, meaning it is completely interpretable without the original text and other claims. Each claim is verified independently, and the results are combined to assess the overall veracity of the original input.\\
\\
Your task is to decompose an input text into exactly \{num\_sub\_claims\} claims while preserving the original meaning. Each claim should be *self-contained*, meaning it is completely interpretable without the original text and other claims.\\
\\
Please directly provide the decomposed set of claims in the following format, each claim should be enclosed with triple backticks:\\
\\
\#\#\# Claims\\
```\\
claim \#1\\
```\\
\\
```\\
claim \#2
```\\
...\\
```\\
claim \#\{num\_sub\_claims\}\\
```\\
\\
Now, please help me decompose the following input text:\\
\\
\#\#\# Input Text\\
```\\
\{input\_text\}\\
```}

\subsection{Decontextualization}
\label{appendix:prompt_decontext}
\texttt{
Definition of Decontextualization: taking a sentence or response together with its context and rewriting it to be interpretable out of context, while preserving its meaning.\\
\\
For question-response pairs, the question is the context, if an response cannot be understood without the context, then the response should be rewrite to incorporate the question.\\
\\
Decontextualize the following response according to the given question, make sure to keep the original meaning of the response. Return the decontextualized response starting with \#\#\# Decontextualized Response:'.\\
\\
\#\#\# Question: \\
\{question\}\\
\\
\#\#\# Response: \\
\{response\}}

\subsection{Decomposition Prompt Instructions}
\label{appendix:prompt_decomp_methods}
\begin{table*}[!htbp]
    \centering
    \begin{tabularx}{\textwidth}{|l|X|}
        \hline
        \textbf{Method} & \textbf{Instruction} \\
        \hline
        VeriScore & You are trying to verify how factual a piece of text is. To do so, you need to break down a sentence and extract as many fine-grained facts mentioned in the sentence as possible. Each of these fine-grained facts should be verifiable against reliable external world knowledge...\\
        \hline
        FactScore & Please breakdown the following sentence into independent facts...\\
        \hline
        Wice & Segment the following sentence into individual facts...\\
        \hline
    \end{tabularx}
    \caption{Partial instructions from different decomposition methods.}
\end{table*}

\end{document}